%% file: 0_main.tex
\documentclass[fleqn,10pt]{wlscirep}
\usepackage[utf8]{inputenc}
\usepackage[T1]{fontenc}
\title{Peer attention enhances student learning}

\author[1,*]{Songlin Xu}
\author[2]{Dongyin Hu}
\author[3]{Ru Wang}
\author[1]{Xinyu Zhang}
\affil[1]{University of California San Diego}
\affil[2]{University of Pennsylvania}
\affil[3]{University of Wisconsin–Madison}

\affil[*]{soxu@ucsd.edu}

\begin{abstract}
Human visual attention is susceptible to social influences. In education, peer effects impact student learning, but their precise role in modulating attention remains unclear. Our experiment (N=311) demonstrates that displaying peer visual attention regions when students watch online course videos enhances their focus and engagement. However, students retain adaptability in following peer attention cues. Overall, guided peer attention improves learning experiences and outcomes. These findings elucidate how peer visual attention shapes students' gaze patterns, deepening understanding of peer influence on learning. They also offer insights into designing adaptive online learning interventions leveraging peer attention modelling to optimize student attentiveness and success.
\end{abstract}
\begin{document}

\flushbottom
\maketitle
%
%
\thispagestyle{empty}


\input{1_intro}

\input{2_result}

\input{3_discuss}

\input{4_method}

\bibliography{sample}

\section*{Acknowledgements}

This work is supported by National Science Foundation Grant NSF CNS-1901048.

\section*{Author contributions statement}

S.X., D.H., R.W., and X.Z. designed research; S.X., D.H., R.W., and X.Z. performed research; S.X., D.H., and X.Z. analyzed data; and S.X., D.H., and X.Z. wrote the paper. All authors have reviewed the paper.

\section*{Additional information}

\textbf{Competing interests}: The authors declare no competing interest.

\end{document}

%% file: 1_intro.tex
\section*{Introduction}
The
attention of humans is profoundly influenced by social cues, a tendency firmly rooted in our nature as inherently social animals \cite{ziman2023predicting,friesen1998eyes,frischen2007gaze,capozzi2018attention,dalmaso2020social,calder2002reading,itier2009neural,graziano2011human,kobayashi1997unique,xu2023tactile}. In educational settings, peer effects readily shape students' attentional patterns, as students naturally follow the gaze of classmates during lessons \cite{yang2014peer}. For instance, when peers orient towards an unexpected disruption, like a bird at the window, others rapidly align to observe the same stimulus. Peer attention impacts learning through multiple pathways - peers can provide information, but students may also conform to peer attentional norms to garner acceptance \cite{asterhan2017teenage,buhs2006peer}.


The existence and nature of peer effects in education have been debated in prior scholarship \cite{burke2013classroom,calvo2009peer,sacerdote2011peer,epple2011peer,winston2004peer,foster2006s,burke2013classroom,smith2009peer,ding2007peers,hanushek2003does}. Some studies reveal positive peer influence on academic achievement \cite{hanushek2003does, ding2007peers}, and peer discussion can enhance understanding even without prior knowledge \cite{smith2009peer}. Peer effects extend to social outcomes as well \cite{sacerdote2011peer}. However, other work finds limited to no linear peer influence on learning, with residential proximity and friendships conferring little added benefit over random peer pairings \cite{foster2006s,burke2013classroom}. While the precise mechanisms continue to be debated, peer effects appear contingent on contextual factors. Overall, prior research confirms peers shape students’ academic experiences, though the specific nature and direction of these influences likely depend on individual, classroom, and school characteristics.


In summary, despite extensive research, peer effects on student learning remain debated, with limited definitive evidence regarding their efficacy. Moreover, granular analysis of real-time student learning behaviors is lacking, which could elucidate mechanisms underlying peer influence. Our research helps address these gaps through a fine-grained investigation correlating learning outcomes with student visual attention patterns under peer influence. Specifically, we test whether peer effects on gaze patterns also impact downstream learning experiences and performance.



Drawing on prior work showing peer cues shape visual attention \cite{downing2004does,frischen2007gaze}, we hypothesize that manipulating peer attention can guide students' eye movements, increase focus, and ultimately improve learning outcomes.
To test this, we conducted an experiment with 311 participants who watched course videos with or without visualized peer attention regions. Fine-grained gaze data revealed participants' attentional engagement and course following behaviors. Regression models decoded relationships between manipulated gaze patterns, attendant learning experiences (e.g., confusion, valid focus, course following), and downstream learning outcomes.


Our results reveal two surprising findings. First, although peer attention manipulated students' gaze, individuals adapted their viewing strategies rather than always mirroring peer focus. For instance, when peer attention diverged from the lecture pace, students' gazes often did not follow. Nevertheless, peer attention cues still increased focus and course following compared to controls.
Second, intentionally guiding students' gaze along the lecture pace did not always improve learning outcomes. Instead, students able to adaptively adjust their focus based on personal needs showed enhanced performance.

%% file: 2_result.tex
\section*{Results}

\subsection*{Task and participants}

\begin{figure*}
\centering
\includegraphics[width=1\linewidth]{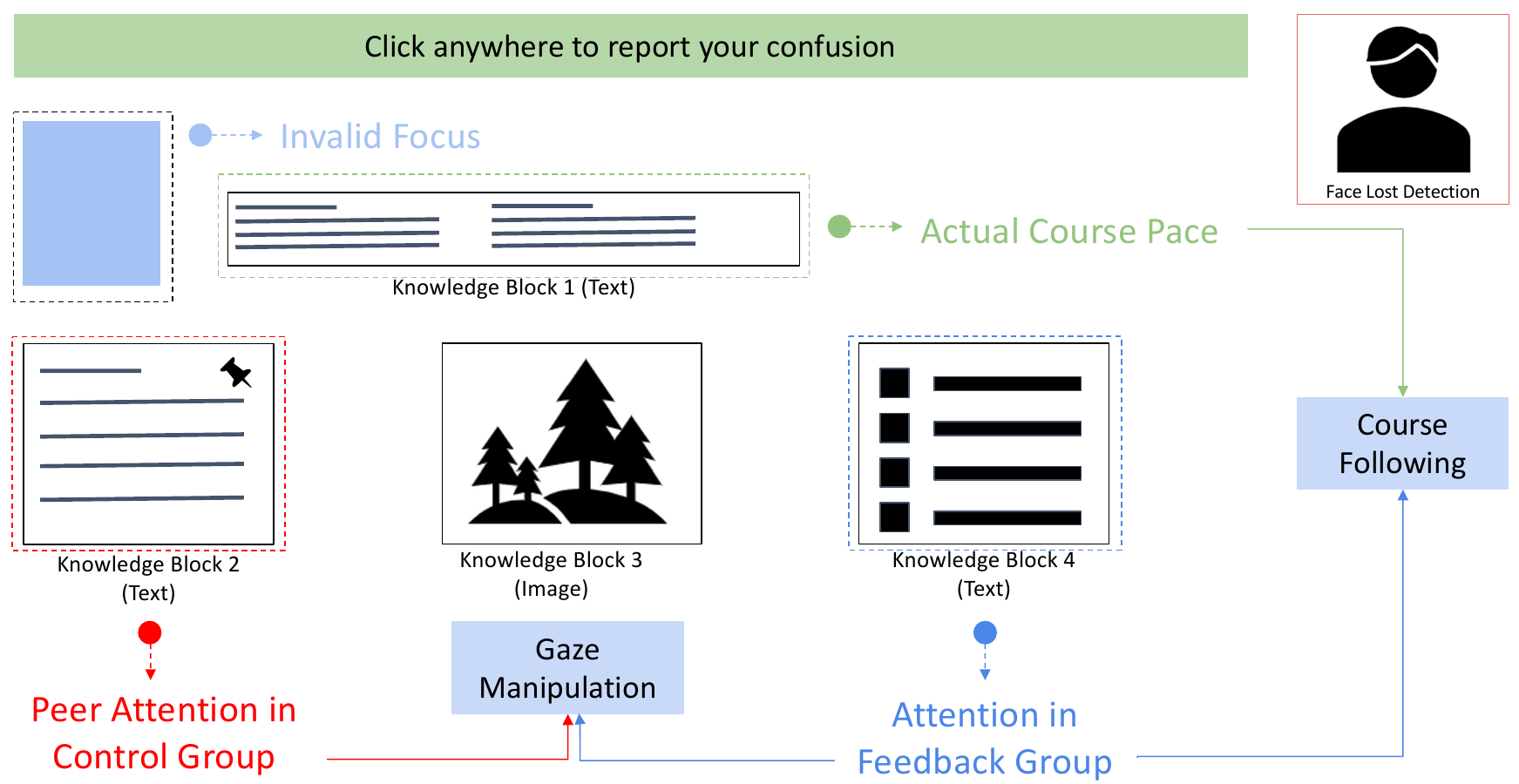}
\caption{Demonstration of metrics in the course video. Knowledge blocks represent meaningful image/text in the course video. If participants' gaze falls into blank areas instead of these knowledge blocks, then it is detected as invalid focus. Course following is measured by comparing the course pace and participants' current gaze focus. Inattention is recorded when face-lost is detected by our system. Confusion is recorded by participants' clicking behaviors on the video. }
\label{f1}
\end{figure*}

We conducted a between-subjects experiment with a Group (Control vs. Feedback) x Video design. Participants were randomly assigned to view one online course video with or without visualized peer attention regions (bounding boxes indicating peer gaze on the slide; see Fig. \ref{f1} and Materials and Methods). Gaze data were collected during viewing. After the video, participants completed a learning assessment and questionnaires. Peer attention regions were extracted from control participants’ gazes. Participants were 311 adults (demographics in Appendix). To isolate effects of the attention manipulation and eliminate effects of course materials, each metric was normalized within each video before ANOVA analysis.

\subsection*{Learning experience}


Analyses of learning experiences revealed significant reductions in inattention duration for Feedback group (59.8\% decrease vs. Controls; $F_{1,301}=6.925, P = 0.009 < 0.01$). Feedback also decreased confusion duration (81.3\% reduction; $F_{1,299}=6.384, P = 0.012 < 0.05$). However, we did not find significant difference on cognitive workload (measured by NASA TLX \cite{NASATLX}) between groups (Feedback: mean $\pm$ SD: 3.355 $\pm$ 0.995; Control: 3.171 $\pm$ 0.958; $F_{1,301}=1.884, P = 0.17 > 0.05$). 61.3\% of Feedback participants also reported noticeable awareness of peer attention cues.

\subsection*{Learning outcome}

Learning outcomes were evaluated using post-video test accuracy. On easy recognition questions (recognition questions whose answers are directly listed in the course video), performance did not significantly differ between groups ($F_{1,301}=3.303, P = 0.07 > 0.05$). However, on hard  questions (comprehensive questions whose answers are not simply listed in the course video), the Feedback group showed significantly higher accuracy versus Controls (19.2\% increase; $F_{1,301}=16.086, P < 0.001$; Feedback: 67.9 $\pm$ 24.8\%, Control: 56.9 $\pm$ 24.3\%). For the overall test, the Feedback group also performed significantly better (10.4\% increase; $F_{1,301}=10.862, P = 0.001$; Feedback: 69.6 $\pm$ 15.7\%, Control: 63.0 $\pm$ 17.8\%) (Fig. \ref{f2}(right)).

\subsection*{Gaze manipulation}

\begin{figure*}
\centering
\includegraphics[width=1\linewidth]{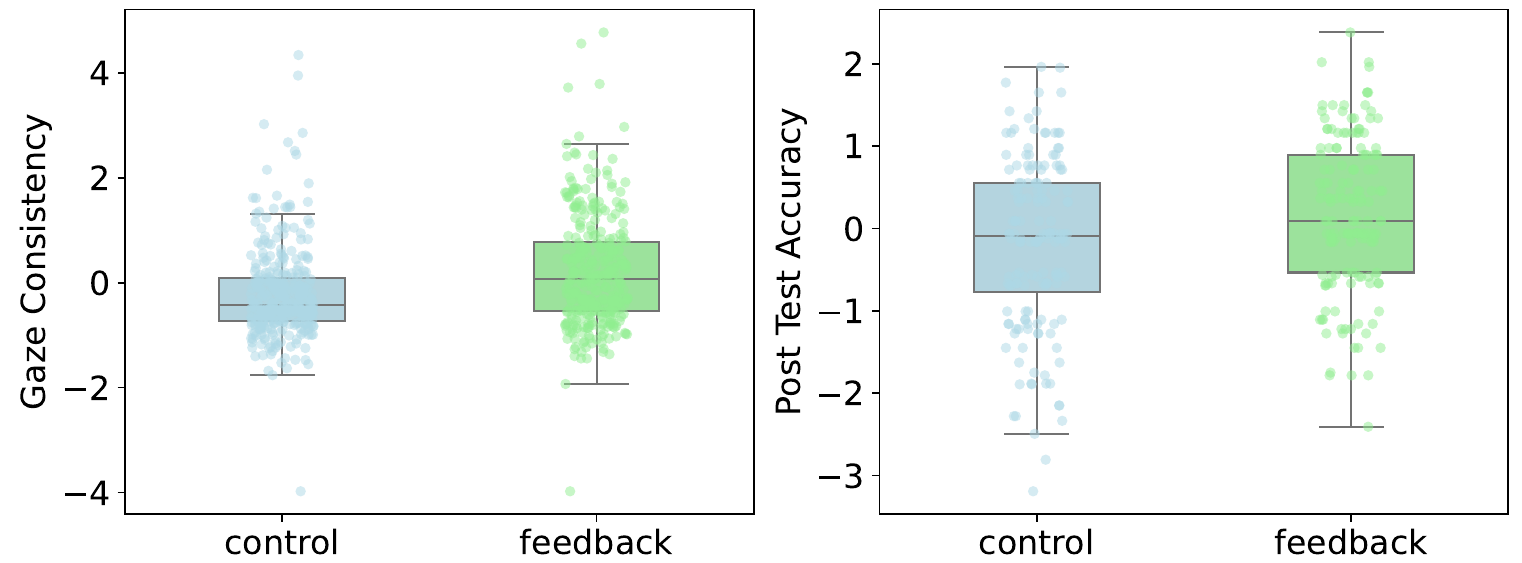}
\caption{Left: Crowd gaze consistency which measures the percentage of students whose gaze falls into similar AoIs in the course slides shared by students in each group. Right: Average accuracy of post test score in each group.}
\label{f2}
\end{figure*}

Analyses of gaze behaviors showed peer attention cues increased valid focus (students' gaze fall into meaningful text/image blocks instead of blank areas in the course video) (8.85\% increase; $F_{1,301}=2.905, P = 0.089$). Feedback also improved course following (12.6\% increase in gaze-pace alignment). As expected, gazes fell more into designated peer attention areas for the Feedback group (12.3\% increase vs. Controls). Critically, the Feedback group showed significantly higher gaze consistency on the same text/image elements ($F_{1,636}=31.402, P < 0.001$) (Fig. \ref{f2}(left)). This revealed that participants in Feedback group exhibited more engagement since crowd gaze consistency could indicate student engagement \cite{madsen2021synchronized}. However, consistent gaze regions differed between groups, suggesting students retained their own strategies to decide whether following peer attention visual region in the course.

\subsection*{Decoding learning process}


In-depth analyses explored correlations between learning experiences and outcomes using Pearson coefficients. As expected, confusion and inattention negatively correlated with group differences and test accuracy on all question types (Fig. \ref{f3}). Valid focus, course following, and gaze focusing positively correlated with group differences and test performance. Peer attention positively influenced test accuracy, with the strongest effect for hard questions.
Interestingly, cognitive workload also positively correlated with group differences and test accuracy. Despite potentially increasing workload, peer attention still improved learning.


We also conducted a granular analysis decoding learning experiences in specific course duration related to each test question (Fig. \ref{f4}). Since question accuracy is binary, logistic regression \cite{cokluk2010logistic,midi2010collinearity} explored correlations with concurrent experiences. Peer attention (coefficient $\beta = 0.84, P < 0.001$) and valid focus (coefficient $\beta = 0.11, P = 0.009 < 0.01$) positively predicted accuracy. However, course following negatively correlated with accuracy (coefficient $\beta = -0.18, P < 0.001$). This divergence between valid focus and course following suggests good pace-following does not guarantee learning. For instance, students pausing to think deeply may exhibit valid focus without strict pace-alignment, yet still understand content.

%% file: 3_discuss.tex
\section*{Discussion}

\begin{figure}
\centering
\includegraphics[width=0.6\linewidth]{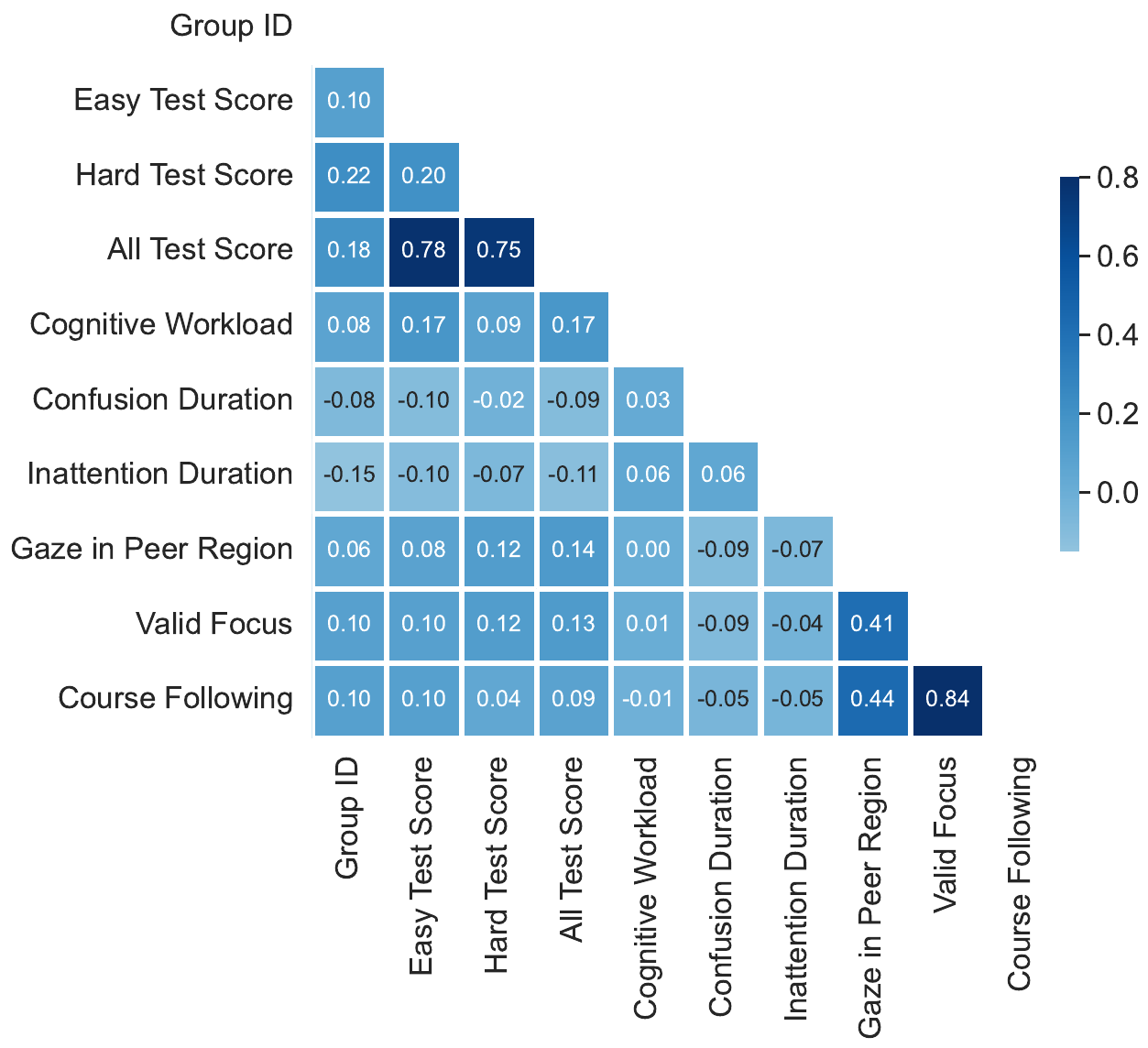}
\caption{Pearson correlation matrix that shows the correlation among different metrics.}
\label{f3}
\end{figure}



While peer effects on student learning have been extensively debated, prior work lacks fine-grained analysis of real-time peer influence on learning processes \cite{calvo2009peer,sacerdote2011peer,epple2011peer,winston2004peer,foster2006s,burke2013classroom,smith2009peer,ding2007peers,hanushek2003does}. Our large-scale experiment helps address this gap through granular gaze and cognitive state data, unveiling how peer visual attention shapes students’ attentional engagement, comprehension, and performance. These insights provide important groundtruth on peer effect mechanisms to inform the design of adaptive educational interventions leveraging peer influence cues to optimize learning experiences and outcomes \cite{kizilcec2020scaling,gielen2010improving,er2021collaborative,xu2023leveraging}.



Our fine-grained analyses reveal several specific mechanisms underlying peer influence on learning. Peer visual cues significantly reduced student inattention and confusion, likely resulting from enhanced attentional focus \cite{felmlee1985peer}. Despite potentially increasing cognitive workload, we did not find significant impact of peer attention on cognitive workload of Feedback group. Over 60\% of students also explicitly noticed peer attention cues, confirming the manipulation's efficacy. However, this does not indicate that peer attention was invalid on other 40\% of students since students' learning behaviors may get influenced unconsciously, suggested by results from the gaze behaviors and learning outcomes.



Peer attention predominantly improved performance on comprehension-based hard questions rather than simple recall easy items. Hard question accuracy showed greater correlation with peer cues versus easy accuracy. This aligns with hard questions requiring sustained attention and deeper conceptual understanding, while easy items can be answered via sporadic focus and recognition \cite{meyer1997computational,reinhard2009need}. Thus, peer attention's benefits manifest strongly for difficult content necessitating engaged focus and mental modeling. Reduced inattention and confusion in the peer attention group further support this explanation - peer cues enhanced attentional regulation and understanding precisely when needed for hard content.


The Feedback group showed significantly higher gaze consistency on shared areas of interest, indicating greater engagement under peer influence \cite{madsen2021synchronized}. However, consistent gazes differed from control viewers, with peer-cued attention better aligned to meaningful regions. This divergence suggests students retain agency in adapting gazes rather than strictly mirroring peers. Despite manipulated gazes, individuals strategically focused based on personal comprehension needs.



Additional analyses further elucidated learning mechanisms. At a coarse-grained level, test accuracy correlated positively with peer attention across question types, affirming overall learning benefits. Peer cues also correlated positively with workload yet negatively with inattention/confusion, suggesting attentional enhancement and confusion reduction underlie effects \cite{felmlee1985peer}. Moreover, manipulated gazes, valid focus, and course following correlated positively with peer attention.
\textit{In summary, decoding the learning process revealed cascading mechanisms - peer cues manipulated gaze behaviors, improving attentional engagement and reducing confusion. This in turn enhanced learning comprehension, evidenced by higher assessment accuracy}. Our multi-level analyses unravel specific processes by which peer visual attention ultimately improves learning outcomes.


Surprisingly, granular analysis in logistic regression revealed negative correlation between course following and accuracy \cite{bill2021following}. One explanation is that multiple factors affect learning, thus pace-adherence alone does not guarantee comprehension. Moreover, eye movements may not perfectly reflect cognition \cite{richardson2005looking} - students could gaze along without understanding.
The divergence between valid focus and course following suggests focused personal thinking, at the cost of pace-alignment, can still enhance understanding. Students pausing the lecture to reflect deeply exhibit valid focus without strict following, yet gain knowledge through mindful comprehension.
Additionally, course following correlated more strongly with easy versus hard question accuracy. Pace-alignment suffices for simple recognition, but complex conceptual learning requires reflective focus beyond passive following.



Our findings advance understanding of peer attention effects on student learning \cite{burke2013classroom,calvo2009peer,sacerdote2011peer,epple2011peer,winston2004peer,foster2006s,burke2013classroom,smith2009peer,ding2007peers,hanushek2003does} in several key ways. We demonstrate differential peer influence under varying task difficulty levels. Granular analysis reveals surprising negative correlation between course following and performance when decoding the learning process. Divergent relationships between experiences (confusion, focus, pace-adherence) and outcomes highlight active attentional adaptation by students rather than passive mirroring of peer cues. These novel insights emerge through large-scale experimentation combined with fine-grained gaze and performance data. Our multifaceted results moves beyond existing literature to elucidate specific mechanisms and boundary conditions of peer effects on attention, comprehension, and achievement. This groundtruth provides directions for effectively leveraging peer-guided visual engagement to enhance student success.

One limitation involves quantifying cognitive states \cite{10.1145/3544548.3580905,xu2023modeling} like inattention and confusion \cite{folstein1975mini}. We used face loss to detect inattention, but this may miss mind wandering without gaze diversion \cite{el2007students}. Similarly, self-reported confusion could miss unreported experiences. More granular sensing like EEG could better capture states, but faces feasibility challenges in large-scale studies. Our webcam and browser events enabled scalable data collection for initial exploration.
Due to self-reports and face loss proxies, inattention and confusion metrics showed high variance and short durations. Some participants never self-reported confusion or lost face tracking. Incorporating gaze metrics like valid focus and course following provides additional attentional engagement evidence. Future work could explore fusing face loss with gaze patterns or facial expressions to better detect mind wandering.


While the overall analysis found significant accuracy differences between groups, results were not always significant when analyzing videos individually. However, group differences remained present in mean values per video. The lack of significance may stem from insufficient sample sizes for individual video analyses. Normalizing across videos to isolate group effects provided adequate power to reveal overall significant differences.
As course material differences were not the focus, we normalized metrics to emphasize peer attention effects over content-specific influences. However, future work should explore peer effects within specific educational contexts using larger samples for video-level analyses.



In conclusion, our large-scale experiment provides novel evidence on peer attention effects in education through fine-grained analysis of learning processes. We demonstrate peer visual cues enhance student engagement, comprehension, and performance via manipulated gaze patterns and reduced mind wandering. Explanatory decoding further unravels specific mechanisms linking peer attention to learning experiences and outcomes. These insights contribute new theoretical grounding to advance the design of adaptive educational interventions \cite{kizilcec2020scaling,gielen2010improving,er2021collaborative} leveraging peer influence. Modelling and guiding visual attention may help optimize student focus, interest, and understanding. Overall, this work elucidates practical pathways for leveraging peer effects, mapped through granular gaze-cognition relationships, to create dynamic learning systems that promote attentiveness and achievement.

%% file: 4_method.tex
\section*{Methods}

\subsection*{Subjects}
We totally recruited 401 adult participants from Prolific \cite{Prolific}. Each volunteer received 2 USD compensation for the 10-min participation. 90 participants did not finish the study completely or gave invalid study data, leaving 311 participants (age: 29.3 ± 6.3 years; 148 females) for analysis.
We used a between-subject Group (Control v.s. Feedback) $\times$ Video (1,2,3,4,5) design where each participant was assigned to one group and only watched one video. For each video, we first recruited 40 participants to finish the study in Control group and then recruited another 40 participants to finish the study in Feedback group. In this way, participants in Feedback group received peer attention feedback which is extracted from the gaze data in the Control group. 
The University of California San Diego Institutional Review Board approved the study, and we obtained written informed consent from all participants beforehand.

\begin{figure*}
\centering
\includegraphics[width=1\linewidth]{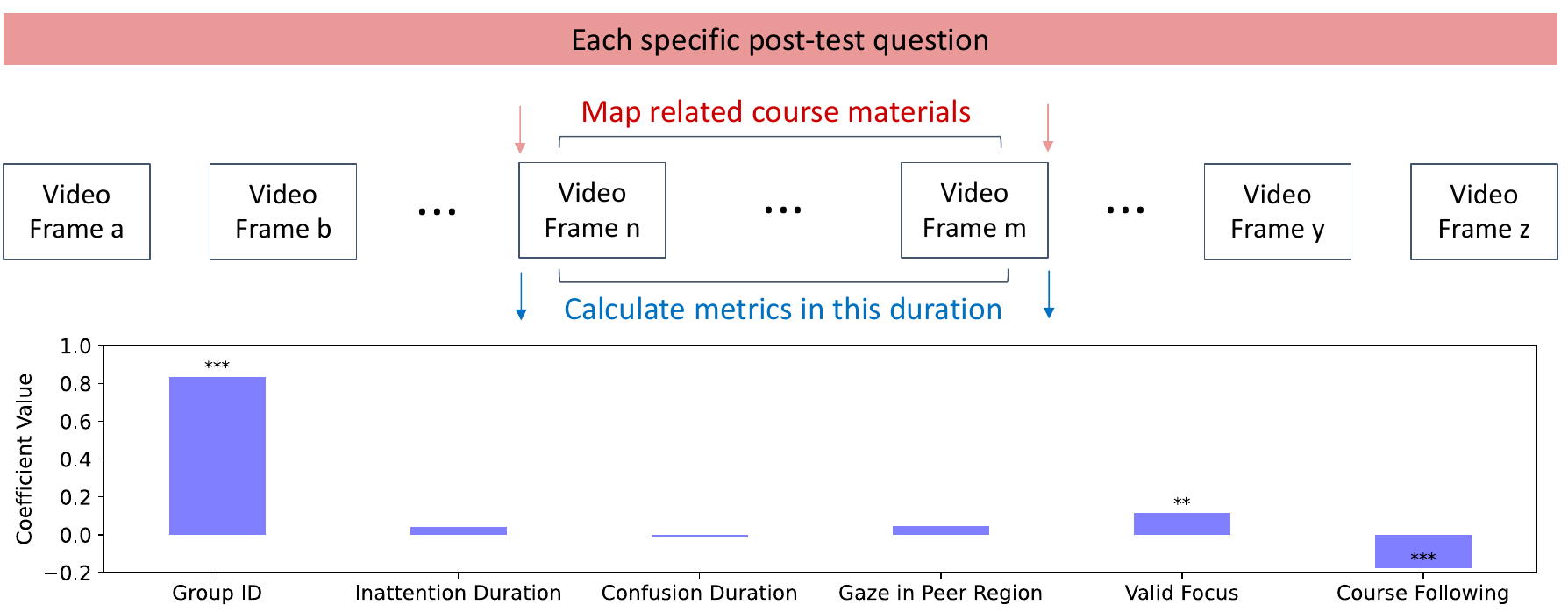}
\caption{Learning decoding process for individual post-test questions. For each question, we first found the related course materials in the video and extracted the timestamps and duration. We then calculated metrics in this specific duration and performed logistic regression to find the correlation between individual question accuracy and metrics.}
\label{f4}
\end{figure*}

\subsection*{Experiment, procedure, course materials}

All participants first read and signed the consent form. Then they accessed a website named CogTeach (our experiment platform). The website introduced the study procedure to participants. Participants performed eye tracking calibration and then watched the corresponding course video. All five course videos are about the topic of AI. Each video duration is around 5 minutes. The specific theme from video 1 to video 5 is \textit{What is AI}, \textit{Supervised Learning}, \textit{Unsupervised Learning}, \textit{Computer Vision}, \textit{AI and Privacy}. To avoid result bias introduced by participants' previous knowledge background, we did not recruit participants who had strong prior knowledge about our course materials.
We have released the video transcripts in the appendix. 
After that, participants finished a post-test including 11-12 questions. Each question and answers are depicted in Appendix. After the post-test, participants were also asked to finish subjective questionnaires. For Control group, the questionnaire just included all questions in NASA TLX form \cite{NASATLX}. For Feedback group, the questionnaire included both NASA TLX form and 8 additional questions about participants' feelings about the peer attention feedback, listed in Appendix Table S2.

\subsection*{CogTeach system}

We developed an experimental online education platform named CogTeach, a browser-based integral video conferencing system that simultaneously enables interaction, cognitive data collection and real-time visual and auditory feedback. 
User data, including gaze, cognitive states, and other interactions, are collected and uploaded to servers, which summarize the information and generate real-time visual or auditory feedback on the client-side. 
The system uses backends implemented in JavaScript and Python. We adopt Flask \cite{Flask} as web service frameworks. Python servers host all algorithms. JavaScript servers serve static files and scripts and manage socket connections for all participants.

\subsubsection*{Gaze collection}

We adopted WebGazer \cite{papoutsaki2016webgazer}, an open-source JavaScript library for eye tracking using web cameras. Instead of uploading video data to a server, WebGazer.js runs entirely on the client-side and preserves user privacy better. 
We adopt a standard 9-point calibration procedure, where users are asked to click points appearing on corners, middle points of each edge, and the center point five times. During the interaction with the web page, mouse clicks are fed to WebGazer for incremental training.

\subsubsection*{Cognitive states}
CogTeach records inattention and confusion states as cognitive measurements.
Student inattention is detected when the gaze estimator detects a missing face from the web camera.
For confusion, students can click on the slide to directly notify the system about which parts they are confused. The system will record the mouse click timestamp and location on the screen for further confusion analysis.

\subsubsection*{Areas of interest (AoIs) detection}
AoIs are detected based on the salient regions of a slide screenshot. As the slides are artificial images created by educators, it usually contains regions that are well separated. We use a hierarchical algorithm based on morphological operations to detect salient regions from a given slide. The input screenshot is initially converted to grayscale, inverted, and resized. The size of the structuring element decreases by a step starting from a given value for each iteration. In each iteration, we use dilation to fill the white space between characters or inside an image and then apply Otsu's method to binarize the dilated image. In this binary figure, we define salient regions as convex hulls of connected components. 
To filter out fragments and overly large regions, this algorithm accepts two tunable parameters to specify the range of the area of regions. Regions with an area within the range are recorded and removed from the image to avoid further divided into smaller regions. The image with regions removed is used in the next iteration.
In the CogTeach system, all screenshots are resized to $960\times 540$, and operations adopts a flat rectangular element that starts with a sidelength of 20 and ends at 5. Regions with an area that falls between 15\% and 1\% of the area of the slide are accepted. 

\subsubsection*{Fixation and saccades detection}

CogTeach system uses a velocity-based method modified from the algorithm proposed in \cite{madariaga2023safide} to detect fixations from the stream of gaze points uploaded from users. Fixations represent a spatial cluster of eye gaze within a time window. Saccades, defined as the movement between fixations, can be identified by velocities. Raw gaze points with speed exceeding a threshold are classified as saccade points, and all other points that fall between two saccades are considered fixations. The horizontal and vertical thresholds $\mu_{x, y}$ are calculated by:
\begin{align}
\mu_{x, y} = \lambda\left(\langle v_{x, y}^2 \rangle - \langle v_{x, y}\rangle^2\right) \label{eq:ek-threshold},
\end{align}
where $\lambda$ is a constant and in our system we use 6, $\langle\cdot\rangle$ denotes the median operator, $v_{x, y}$ are vectors of horizontal and vertical speeds.

The original algorithm applies to a batch of gaze points. Since our system offers real-time processing, the original algorithm is modified to be applied over the gaze points in a certain time window. We add a few more changes, such as filters to smooth the gaze points, limiting the minimal fixation duration ($200$ ms), and removing blinks.

\subsubsection*{Gaze clustering}
Fixations are aligned with AoIs detected using the method described above. Each detected fixation is aligned to the closest AoI detected from the screenshot. The distance measurement is defined as the minimum distance between the fixation center and each line segment of the convex hull. The fixation center is the mean of all gaze points belonging to the fixation.
In addition, confusion reported by the students are aligned with AoIs in the same way. Thus, each AoI is associated with a few fixation points and confusion reports.

\subsubsection*{Peer attention region visualization}

To visualize peer attention region, we extracted fixation and cluster user gaze from Control group. By comparing these gaze clusters and AoIs extracted from salient regions of slides, we vote one AoI shared by most users in Control group in every 5 seconds, which is then presented to participants in Feedback group, named as peer attention feedback (Fig. \ref{f1}). Such voted AoI is in the form of a circumscribed red rectangle to replace the convex hulls to reduce the complexity and cognitive load for understanding.

\subsection*{Evaluation metrics}
For learning experience, we used inattention duration, confusion duration, cognitive workload, and feedback experience for evaluation. The inattention duration is calculated by the duration when face lost is detected in CogTeach system. The confusion duration is calculated by the duration when users click on slides to report their confusion. Cognitive workload is measured by NASA TLX form \cite{NASATLX} where we ask participants to rate their mental demand, physical demand, and so on (details in Appendix Table S1). Feedback experience is evaluated by a  questionnaire to ask users to rate their subjective feelings about the peer attention feedback (details in Appendix Table S2).

For learning outcome, we used average accuracy of easy questions, hard questions, and all questions in post-test for evaluation. Easy questions refer to those recognition questions whose answers are just listed in the course video. So students could easily answer these questions as long as they have watched the video. Hard questions refer to the comprehensive questions whose answers are not simply listed in the course video. Students need to not only watch the video but also understand the concepts well in the course in order to answer them correctly. There are totally 11-12 questions in the post-test for each course video. Question details are listed in Appendix.

For gaze manipulation, we used valid focus ratio, course following ratio, gaze in peer region, and crowd AoI gaze consistency for evaluation. Students' focus is considered to be valid if their gaze falls into meaningful AoIs that include text/image blocks instead of blank areas in the course video. Course following is measured by comparing the AoIs in the current course pace and AoIs that students are currently paying attention to. For example, if the course is talking about contents in some specific AoIs but students are focusing on other AoIs, then the students are not following the course pace. We also measure the student gaze falling into the peer attention visual region, which is the AoI voted by most participants in the Control group. Crowd AoI gaze consistency measures the 
percentage of students whose gaze falls into similar AoIs in the course slides shared by students in each group. Larger crowd AoI consistency indicates more consistent student focus on the course contents in the crowd.

\subsection*{Data and Materials Availability}

All data and codes are publicly available without restriction: https://github.com/songlinxu/PeerEdu.